# Room Temperature Terahertz Spectrometer with Quantum-Level Sensitivity

Ning Wang[1,2] †, Semih Cakmakyapan[1] †, Yen-Ju Lin[1], Hamid Javadi[3], Mona Jarrahi[1,2]*

[1]Electrical and Computer Engineering Department, University of California Los Angeles, USA.
[2]Electrical Engineering and Computer Science Department, University of Michigan, Ann Arbor, USA.
[3]Jet Propulsion Laboratory, California Institute of Technology, Pasadena, USA.

*Correspondence to: mjarrahi@ucla.edu
†Equal contribution

**Photon detection with quantum-level sensitivity is particularly challenging in the terahertz regime (0.1-10 THz), which contains ~98% of all the photons existing in the universe. Near-quantum-limited terahertz spectrometry has so far only been possible through the use of cryogenically cooled superconducting mixers as frequency downconverters[1,2]. Here we introduce a spectrometry scheme that uses plasmonic photomixing for frequency downconversion to offer quantum-level sensitivities at room temperature for the first time. Frequency downconversion is achieved by mixing terahertz radiation and a heterodyning optical beam with a terahertz beat frequency in a plasmonics-enhanced semiconductor active region. We demonstrate spectrometer sensitivities down to 3 times the quantum-limit at room temperature. With a versatile design capable of broadband spectrometry, this plasmonic photomixer has broad applicability to quantum optics, chemical sensing, biological studies, medical diagnosis, high data-rate communication, as well as astronomy and atmospheric studies.**

Detection of faint fluxes of photons at terahertz frequencies is crucial for various applications[3,4] including biosensing[5-7], medical diagnosis[8,9], chemical detection[10,11], atmospheric studies[12,13], space explorations[14,15], high-data-rate communication[16,17], and security screening[18,19]. Heterodyne terahertz spectrometers based on cryogenically cooled superconducting mixers have so far been the only instruments that can provide high spectral resolution and near-quantum-limited sensitivity levels[1,2]. The operation temperature, bandwidth constraints, and complexity of these terahertz spectrometers have restricted their use to mostly astronomy and atmospheric studies[20-22], limiting the overall impact and wide-spread use of terahertz technologies. The presented terahertz spectrometry scheme is governed by new physical processes that enable, for the first time, quantum-level sensitivity, broad spectral bandwidth, and high spectral resolution while operating at room temperatures.

Fig. 1 provides more insight into the operation of this terahertz spectrometer. It consists of a plasmonic photomixer (i.e., a photomixer with plasmonic contacts) integrated with a logarithmic spiral antenna on a photo-absorbing semiconductor substrate. The plasmonic photomixer is pumped by a heterodyning optical beam with a terahertz beat frequency, $\omega_{beat}$. Therefore, the concentration of the photo-generated carriers follows the intensity envelope of the optical pump beam, $P_{pump}[1+\cos(\omega_{beat}t)]$. When a terahertz radiation at $\omega_{THz}$ is received by the logarithmic spiral antenna a terahertz electric field is induced across the photomixer contact electrodes, $E_{THz}\cos(\omega_{THz}t)$, which drifts the photo-generated carriers. The induced drift photocurrent contains an intermediate frequency (IF) component at $|\omega_{beat} - \omega_{THz}|$, which falls in the radio frequency (RF) range by appropriately selecting the pump beat frequency. The induced IF signal is then routed through a low-noise amplifier and a bandpass filter and is detected by an RF power detector, as illustrated in Fig. 1a. The detected RF power carries the received spectral information at $\omega_{beat} \pm \omega_{BPF}$ over a spectral bandwidth equal to the bandwidth of the bandpass filter. By tuning the optical pump beat frequency and recording the detected power, the received terahertz spectrum is extracted over a broad frequency range determined by the logarithmic spiral antenna bandwidth. Moreover, the received terahertz spectrum is resolved with a high spectral resolution determined by the bandwidth of the bandpass filter and the linewidth of the optical pump beam. The described spectrometer operates in a double sideband (DSB) mode of observation, where the spectral information of the desired terahertz frequency at $\omega_{beat} - \omega_{BPF}$ is measured together with that of the image frequency, $\omega_{beat} + \omega_{BPF}$. For a single sideband (SSB) operation, a quasi-optical diplexer can be placed along the received radiation path to filter out the image frequency[23].

The presented terahertz spectrometry scheme offers transformative advantages over conventional techniques, which utilize a Schottky diode, hot electron bolometer (HEB), or superconductor-insulator-superconductor (SIS) mixer together with a terahertz local oscillator for THz-to-RF downconversion[24-27]. While the conversion gain of conventional mixers is limited by the low radiation power of existing terahertz local oscillators, significantly higher conversion gains are offered by plasmonic photomixers because of the availability of high-power optical pump sources. The presented plasmonic photomixer is designed such that the IF output and noise powers have a quadratic and linear dependence on the optical pump power, respectively. As a result, the spectrometer signal-to-noise ratio (SNR) and dynamic range can both be increased by boosting the optical pump power. By controlling the photomixer SNR through the optical pump power level, we demonstrate terahertz spectrometry with quantum-level sensitivities without the need for cryogenic cooling, which is required by SIS and HEB mixers to offer similar sensitivities. Another major advantage of the new terahertz spectrometry scheme is that its spectral bandwidth is determined by the wavelength tuning range of the optical pump source. Therefore, it offers broad spectral bandwidths that cannot be achieved by conventional techniques because of the limited tunability of existing terahertz local oscillators. We demonstrate terahertz spectrometry over the 0.1-5 THz frequency range with a spectral resolution and stability of less than 1 kHz through a single plasmonic photomixer. To achieve the same spectral bandwidth through conventional techniques, a large number of cryogenically cooled superconducting mixers and terahertz local oscillators operating over the 0.1-5 THz frequency range would be required to offer similar sensitivities.

Figure 1b shows a scanning electron microscopy image of a fabricated plasmonic photomixer prototype (see Methods). A short carrier lifetime semiconductor is used as the photo-absorbing substrate (i.e., a low-temperature-grown GaAs substrate with a 0.3 ps carrier lifetime) to recombine the slow photocarriers that degrade the THz-to-RF conversion efficiency. To achieve a broad spectral bandwidth, the geometry of the logarithmic spiral antenna is chosen to offer a radiation resistance of ~70 Ω and a negligible reactance over the 0.1-5 THz frequency range. To achieve high THz-to-RF conversion efficiencies, two nanoscale Ti/Au gratings are used as the photomixer contacts, and their dimensions are chosen to enhance the optical pump intensity at the contact-semiconductor interface through the excitation of surface plasmon waves[28], as illustrated in Fig. 1c. As a result, a large fraction of the photo-generated carriers is concentrated in close proximity to these contacts, termed plasmonic contacts. By reducing the average transport path length of the photo-generated carriers to the contact electrodes, a larger number of the photocarriers drift to the contact electrodes in response to an induced terahertz electric field, and higher THz-to-RF conversion efficiencies are achieved.

The operation of the fabricated plasmonic photomixer as a heterodyne terahertz spectrometer is first characterized at 0.55 THz (see Methods and Fig. S3a). Two wavelength-tunable, distributed-feedback (DFB) lasers are used to provide the heterodyning optical pump beam. Fig. 2a shows the observed IF spectrum centered at 1 GHz (blue curve) and the noise/background spectrum in the absence of the terahertz radiation (red curve). The lower spectral peaks observed near 750 MHz and 900 MHz are the background radio signals picked up by the IF transmission lines and cables. The observed IF signal has a linewidth of ~3 MHz, which is dominated by the linewidth of the two DFB lasers that provide the heterodyning optical pump beam. To investigate the impact of the optical pump linewidth on the spectral resolution of the presented spectrometry system, a highly stable optical comb from a Ti:sapphire mode-locked laser is used as the optical pump beam (see Methods and Fig. S3b). The observed spectrum near 1 GHz (Fig. 2a inset) has a linewidth of 1 kHz FWHM, which is dominated by the linewidth of the 0.55 THz source. Therefore, the fabricated plasmonic photomixer enables terahertz spectrometry with resolutions less than 1 kHz, similar to the spectral resolution of conventional terahertz spectrometry systems.

The impact of the optical pump power on the sensitivity and dynamic range of the plasmonic photomixer is analyzed by recording the IF signal and noise power levels at different optical pump powers. Fig. 2b shows the measured IF SNR (the ratio between the IF signal and noise powers) as a function of the optical pump power, indicating a linear increase in the SNR at optical powers below 33 mW and a reduction in the SNR at optical powers above 33 mW. This trend can be explained by different physical mechanisms affecting the photomixer noise and photocarrier dynamics inside the device active area. At low optical pump powers, the device noise is dominated by the Johnson-Nyquist noise[29]. Therefore, the IF noise power, which is inversely proportional to the photomixer resistance, increases linearly as a function of the optical pump power. The laser noise becomes more dominant at high optical pump powers, increasing the rate of the IF noise growth as a function of the optical pump power, as illustrated in Fig. 2c. On the other hand, the IF signal power, which is quadratically proportional to the induced drift photocurrent, increases quadratically as a function of the optical pump power at low optical pump powers. The carrier screening effect becomes more dominant at high photocarrier concentrations, decreasing the rate of the IF signal growth as a function of the optical pump power, as illustrated in Fig. 2d.

The plasmonic photomixer sensitivity is characterized using the Y-factor method, which measures the IF response of the plasmonic photomixer to input noise sources from hot and cold loads (see Methods and Fig. S3c). Unlike most studies that use a single hot/cold load for Y-factor measurements, multiple hot/cold loads are used in our sensitivity analysis to simultaneously assess the spectrometer linearity. Figs. 3a and 3b show the measured IF power in response to thermal loads varying between 77 K and 1500 K at a 30 mW optical pump power. The observed roll-off in the IF power at higher frequencies is due to the plasmonic photomixer parasitics and antenna frequency response (Fig. 3a). As expected, a linear relation between the IF power and the load temperature is observed over the entire 77-1500 K range (Fig. 3b). The Y-factor and DSB noise temperature values are calculated from the measured IF powers at 1500/295 K hot/cold loads using the Callen-Walton effective hot/cold temperatures[30] (see Methods). Although 1500 K and 295 K are used as the hot and cold loads, the linear dependence of the IF power on the load temperature (Fig. 3b) indicates that the same Y-factor and DSB noise temperature values can be achieved when using other hot/cold temperatures. DSB noise temperatures of 120-700 K are achieved in the 0.1-5 THz frequency range at a 30 mW optical pump power at room temperature (Fig. 3c), corresponding spectrometer sensitivity values down to 3 times the quantum noise limit ($h\nu/2k$).

Figure 3c compares the DSB noise temperature of the fabricated plasmonic photomixer with previously demonstrated Schottky, HEB, and SIS mixers used in conventional heterodyne spectrometers in the 0.3-5 THz frequency range[31]. This comparison indicates the superior performance of the presented plasmonic photomixer in offering noise temperatures lower than the cryogenically cooled HEB mixers and SIS mixers at frequencies above 0.8 THz (down to 3 times that of the quantum noise limit) without the need of a terahertz local oscillator and while operating at room temperature (see Methods). Remarkably, this unprecedented performance is achieved by a single plasmonic photomixer and optical pump beam with a beat frequency tunability of 5 THz. To achieve similar sensitivities and spectral bandwidths through conventional techniques, a large number of cryogenically cooled SIS mixers, HEB mixers, and terahertz local oscillators would be required[20-22]. It should be noted that the sensitivity measurements are performed at frequencies away from strong water absorption lines. By sweeping the beat frequency of the heterodyning optical pump beam and recording the IF power, water absorption lines matching the HITRAN database are observed.

The introduced terahertz spectrometry scheme opens new opportunities at the interface of quantum optics and terahertz photonics and offers a versatile experimental platform for fundamental studies in physics, chemistry, biology, material science, and astronomy. The presented terahertz spectrometer is capable of offering higher detection sensitivities at lower frequencies by using a larger diameter silicon lens, and the demonstrated detection sensitivity levels at lower frequencies is limited by the low radiation coupling to the utilized silicon lens with a 1.2 cm diameter. Additionally, the presented terahertz spectrometer is capable of offering even higher detection sensitivities by using a bandpass filter with a narrower bandwidth, ideally matching the linewidth of the heterodyning optical pump beam (see Methods). The demonstrated sensitivity levels are limited by the 15 MHz bandwidth of the bandpass filter, which is substantially larger than the optical pump linewidth, but chosen to be slightly larger than the DFB laser beat frequency fluctuations. By using higher stability lasers and bandpass filters with narrower bandwidths, the spectrometer noise power level can be considerably reduced without impacting the signal power, thus, higher detection sensitivities can be achieved. Ultimately, the introduced plasmonic photomixer can be integrated with polarization-sensitive antennas (Fig. S4) to determine minute anisotropies in the terahertz radiation polarization.

## Methods

**Photomixer Fabrication.** The plasmonic photomixer is fabricated on a LT-GaAs substrate. The fabrication process starts with patterning two plasmonic contact electrode gratings, by electron beam lithography, followed by Ti/Au (5/45 nm) deposition and liftoff. A 300-nm-thick $Si_3N_4$ anti-reflection coating layer is then deposited using plasma-enhanced chemical vapor deposition. Then, two contact vias are patterned by optical photolithography and etched through the $Si_3N_4$ layer by dry plasma etching. Finally, the logarithmic spiral antenna, IF transmission line, and bonding pads are patterned using optical photolithography, followed by Ti/Au (50/400 nm) deposition and liftoff. The fabricated plasmonic photomixer is mounted on a silicon lens (1.2 cm in diameter) glued onto a printed circuit board (PCB) with an SMA connector. The device output pads are bonded to the leads of the SMA connector to extract the IF output signal. The PCB is placed on a rotation mount to enable optical pump polarization adjustments. The silicon lens is glued to a tapered metallic cylinder (inner/outer diameter of 1.5/4.5 cm and length of 2.5 cm) to assist radiation coupling at wavelengths comparable to or larger than the silicon lens dimensions.

**Numerical Simulations of the Designed Plasmonic Gratings.** A finite-element-method-based electromagnetic software package (COMSOL) is used to characterize the interaction of the optical pump with the designed nanoscale Ti/Au gratings with a 50 nm thickness, 200 nm pitch, 100 nm spacing, and 300 nm thick $Si_3N_4$ anti-reflection coating. Figure S1a shows the power transmission of a *y*-polarized optical pump through the nanoscale gratings into the LT-GaAs substrate, predicting an optical power transmission of 85% at an ~784 nm optical wavelength. Since the optical transmission through the plasmonic gratings is accompanied by the excitation of surface plasmon waves, a large fraction of the photo-generated carriers is concentrated in close proximity to the contact electrode, as illustrated in Fig. S1b.

**Spectrometer Characterization.** The operation of the fabricated plasmonic photomixer as a heterodyne terahertz spectrometer is characterized in response to radiation from a ×2×3 frequency multiplier chain developed at JPL for the Herschel Space Observatory that upconverts the frequency of a Gunn oscillator (Millitech GDM-10 SN224) to 0.55 THz. To provide the heterodyning optical pump beam, the outputs of two wavelength-tunable, distributed-feedback (DFB) lasers with center wavelengths of 783 nm and 785 nm (TOPTICA #LD-0783-0080-DFB-1 and #DLC-DL-PRO-780) are combined and amplified (Toptica BoosTA Pro) to provide a tunable optical beat frequency from 0.1 to 5 THz. The IF output of the plasmonic photomixer is amplified using a low-noise amplifier (Mini-Circuits ZRL-1150) and monitored by an electrical spectrum analyzer. The schematic diagram of the experimental setup is shown in Fig. S3a.

To investigate the impact of the optical pump linewidth on the spectral resolution of the presented spectrometry system, a highly stable optical comb from a Ti:sapphire mode-locked laser with a comb spacing of 78 MHz is used as the optical pump beam. The schematic diagram of the experimental setup is shown in Fig. S3b.

The plasmonic photomixer sensitivity is characterized using the standard Y-factor method, which measures the IF response of the plasmonic photomixer to the input noise sources from hot and cold loads. Room-temperature and liquid-nitrogen-soaked TK Instruments Tessellating THz RAM absorber tiles are used to provide the thermal loads at 295 K and 77 K, respectively. Calibrated blackbody (IR-564 from Boston Electronics) and Globar sources (Thorlabs - SLS203L) are used to provide the thermal loads in the 295-1500 K range. The optical pump beam from the dual DFB laser system is modulated using an acousto-optic modulator (Gooch & Housego AOMO 3080-125) at a 100 kHz rate, and the output IF signal at ~1 GHz is detected by a power meter (Mini-Circuits ZX47-60LN) using a lock-in amplifier with the 100 kHz modulation reference frequency and a 1 s time-constant/integration time (Fig. S2). A low-noise amplifier (Mini-Circuits ZRL-1150) and a bandpass filter (Mini-Circuits ZVBP-909) with a 15 MHz bandwidth are used to further condition the IF signal before the power meter. The measurements are performed in air, without the use of any vacuum or purging. The schematic diagram of the experimental setup is shown in Fig. S3c.

**Spectrometer Noise Temperature Calculation.** The output noise of a heterodyne spectrometer is determined by both the input noise to the spectrometer, $N_i$, and the noise produced by the spectrometer, $N_s$, given by

$$N_s = kT_s B \tag{1}$$

where $k$ is Boltzmann's constant, $B$ is the spectrometer bandwidth, and $T_s$ is the noise temperature of the spectrometer. It should be emphasized that the spectrometer noise temperature, $T_s$, is not the physical temperature of the spectrometer but rather is an equivalent temperature that produces the same amount of noise. As a result, the noise temperature, $T_s$, can be lower than the environmental temperature for a well-designed high-sensitivity spectrometer system. One common way to measure the spectrometer noise temperature is the Y-factor method, which measures the ratio of the spectrometer output to the input noise sources from a hot and cold load

$$Y = \frac{N_s + N_{i-hot}}{N_s + N_{i-cold}} \tag{2}$$

where $N_{i-hot}$ and $N_{i-cold}$ are the input noise powers from the hot and cold loads, respectively. The equivalent input noise power from the spectrometer is calculated from the measured Y-factor

$$N_s = \frac{N_{i-hot} - YN_{i-cold}}{Y-1} \tag{3}$$

where the noise powers from the hot/cold loads are given according to the Callen and Welton law

$$N_i = kTB\left[\frac{\frac{h\nu}{kT}}{\exp\left[\frac{h\nu}{kT}\right]-1}\right] + \frac{h\nu}{2}B \tag{4}$$

where $h$ is Planck's constant, $\nu$ is the photon frequency, $h\nu/2$ is the quantum noise present even at absolute zero temperature, and $T$ is the physical temperature of the thermal load. Therefore, the equivalent input noise temperature of the spectrometer is calculated as

$$T_s = \frac{T_{eff.hot} - YT_{eff.cold}}{Y-1} \tag{5}$$

where $T_{eff.hot}$ and $T_{eff.cold}$ are the equivalent noise temperatures of the hot and cold loads, respectively.

$$T_{eff} = T\left[\frac{\frac{h\nu}{kT}}{\exp\left[\frac{h\nu}{kT}\right]-1}\right] + \frac{h\nu}{2k} \tag{6}$$

**Minimum Number of Detectable Photons.** The output noise of a heterodyne spectrometer when operating at a physical temperature $T_0$ is given by

$$N_0 = (N_i + N_s)G = kT_s BG + kT_0 BG\left[\frac{\frac{h\nu}{kT_0}}{\exp\left[\frac{h\nu}{kT_0}\right]-1}\right] + \frac{h\nu}{2}BG \tag{7}$$

where $G$ is the gain of the spectrometer. On the other hand, the spectrometer output power for a received input signal power of $S_i$ is given by

$$S_0 = GS_i \tag{8}$$

therefore, the signal-to-noise-ratio at the spectrometer output is

$$SNR = \frac{S_0}{N_0} = \frac{S_i}{kT_s B + kT_0 B\left[\frac{\frac{h\nu}{kT_0}}{\exp\left[\frac{h\nu}{kT_0}\right]-1}\right] + \frac{h\nu}{2}B} \tag{9}$$

Assuming a signal-to-noise ratio of 1 is required for the minimum detectable signal, the minimum number of detectable photons is calculated as

$$\frac{kT_s}{h\nu} + \frac{kT_0}{h\nu}\left[\frac{\frac{h\nu}{kT_0}}{\exp\left[\frac{h\nu}{kT_0}\right]-1}\right] + \frac{1}{2} \tag{10}$$

Therefore, the minimum number of detectable photons is determined by both the spectrometer physical temperature and the spectrometer noise temperature (which can be lower than the spectrometer physical temperature). At low terahertz frequencies ($h\nu \ll KT_0$) the minimum number of detectable photons is dominated by the input thermal noise to the spectrometer. However, at higher terahertz frequencies at which the photon energy becomes comparable or larger than the input thermal noise to the spectrometer, the minimum number of detectable photons is dominated by the spectrometer noise temperature. Therefore, single-photon detection sensitivities can be achieved at higher terahertz frequencies. For our fabricated plasmonic photomixer prototype, which offers a DSB noise temperature of ~1.5 $h\nu/k$ at 3 THz, the minimum number of detectable terahertz photons at room temperature (295 K) and 77 K are 4 and 3, respectively. Here we assume the same device noise temperature at 295 K and 77 K. However, since the device noise is dominated by the Johnson-Nyquist noise, lower device noise temperatures and minimum number of detectable terahertz photons are expected at 77 K.

**Linewidth of the Heterodyning Optical Pump.** The spectrum of the heterodyned optical pump beam can be calculated from the spectra of the two DFB lasers forming the optical pump beam. The spectral profiles of the two DFB lasers used in our setup, $f_{DFB1}(\omega)$ and $f_{DFB2}(\omega)$, are modeled by Gaussian functions with center frequencies of $\omega_{DFB1}$ and $\omega_{DFB2}$ and $1/e^2$ linewidths of $4\sigma_{DFB1}$ and $4\sigma_{DFB2}$, respectively.

$$f_{DFB1}(\omega) = \frac{1}{\sqrt{2\pi}\sigma_{DFB1}} \exp\left(-\frac{(\omega-\omega_{DFB1})^2}{2\sigma_{DFB1}^2}\right) \tag{11}$$

$$f_{DFB2}(\omega) = \frac{1}{\sqrt{2\pi}\sigma_{DFB2}} \exp\left(-\frac{(\omega-\omega_{DFB2})^2}{2\sigma_{DFB2}^2}\right) \tag{12}$$

Therefore, the electric field of the heterodyned optical pump beam, which is the superposition of the electric fields of the two DFB laser beams, $E_{DFB1}$ and $E_{DFB2}$, is calculated as

$$E_{pump}(t) = E_{DFB1} + E_{DFB2} = E_0 \int \sqrt{f_{DFB1}(\omega)} e^{j\omega t} d\omega + E_0 \int \sqrt{f_{DFB2}(\omega)} e^{j\omega t} d\omega \tag{13}$$

where $E_0$ is the electric field of the balanced DFB lasers. As a result, the power spectrum of the heterodyned optical pump beam at the beat frequency of the two DFB lasers is calculated as

$$P_{pump}(\omega) = \frac{|E_0|^2}{2\eta_0} \frac{1}{\sqrt{2\pi}\sigma_{pump}} \exp\left(-\frac{(\omega-\omega_{THz})^2}{2\sigma_{pump}^2}\right) \tag{14}$$

where $\eta_0$ is the characteristic impedance of free space, and $\omega_{THz}$ is the angular beat frequency ($\omega_{DFB1} - \omega_{DFB2}$) set to be in the terahertz range. Therefore, the resulting heterodyned optical pump beam has a Gaussian spectrum with a $1/e^2$ linewidth of $4\sigma_{pump}$, where $\sigma_{pump} = (2\sigma_{DFB1}^2 + 2\sigma_{DFB2}^2)^{1/2}$.

**References**

1. Tucker, J. Quantum limited detection in tunnel junction mixers. *IEEE J. Quantum Electron.* **15**, 1234-1258 (1979).
2. Richards, P. L. & Hu, Q. Superconducting components for infrared and millimeter-wave receivers. *Proc. IEEE* **77**, 1233-1246 (1989).
3. Tonouchi, M. Cutting-edge terahertz technology. *Nat. Photon.* **1**, 97-105 (2007).

4. Ho, L., Pepper, M. & Taday, P. Terahertz spectroscopy: Signatures and fingerprints. *Nat. Photon.* **2**, 541-543 (2008).
5. Brucherseifer, M. *et al*. Label-free probing of the binding state of DNA by time-domain terahertz sensing. *Appl. Phys. Lett.* **77**, 4049-4051 (2000).
6. Ogawa, Y. *et al*. Interference terahertz label-free imaging for protein detection on a membrane. *Opt. Exp.* **16**, 22083-22089 (2008).
7. Cheon, H. *et al*. Terahertz molecular resonance of cancer DNA. *Sci. Rep.* **6**, 37103 (2016).
8. Son, J. H. Terahertz biomedical science and technology. CRC Press (2014).
9. Woodward, R. M. *et al*. Terahertz pulse imaging in reflection geometry of human skin cancer and skin tissue. *Phys. Med. & Bio.* **47**, 3853-3863 (2002).
10. Liu, K. *et al*. Characterization of a cage form of the water hexamer. *Nat.* **381**, 501-503 (1996).
11. Jacobsen, R. H., Mittleman, D. M. & Nuss, M. C. Chemical recognition of gases and gas mixtures with terahertz waves. *Opt. Lett.* **21**, 2011-2013 (1996).
12. Manney, G. L. *et al*. Unprecedented Arctic ozone loss in 2011. *Nat.* **478**, 469-475 (2011).
13. Solomon, S. *et al*. Contributions of stratospheric water vapor to decadal changes in the rate of global warming. *Sci.* **327**, 1219-1223 (2010).
14. Neugebauer, G. *et al*. Early results from the infrared astronomical satellite. *Sci.* **224**, 14-21 (1984).
15. Phillips, T. G. & Keene, J. Submillimeter astronomy (heterodyne spectroscopy). *Proc. IEEE*, **80**, 1662-1678 (1992).
16. Kleine-Ostmann, T. & Nagatsuma, T. A review on terahertz communications research. *J. Inf. Mill. & THz Wave.* **32**, 143-171 (2011).
17. Koenig, S. *et. al*. Wireless sub-THz communication system with high data rate. *Nat. Photon.* **7**, 977-981 (2013).
18. Cooper, K. B. *et al.* Penetrating 3-D imaging at 4-and 25-m range using a submillimeter-wave radar. *Trans. Microwave Theory Tech.* **56**, 2771-2778 (2008).
19. Federici, J. F. *et al*. THz imaging and sensing for security applications—explosives, weapons and drugs. *Semiconduct. Sci. Tech.* **20**, S266-S280 (2005).
20. De Graauw, T. *et al*. The Herschel-heterodyne instrument for the far-infrared (HIFI). *Astronomy & Astrophysics* **518**, L6 (2010).
21. Wootten, A. & Thompson, A. R. The Atacama large millimeter/submillimeter array. *Proc. IEEE* **97**, 1463-1471 (2009).
22. Heyminck, S. *et al*. GREAT: the SOFIA high-frequency heterodyne instrument. *Astronomy & Astrophysics* **542**, L1 (2012).
23. Pickett, H. M. & Chiou, A. E. Folded Fabry-Perot quasi-optical ring resonator diplexer: theory and experiment. *Trans. Microwave Theory Tech.* **31**, 373-380 (1983).
24. Wengler, M. J. Submillimeter-wave detection with superconducting tunnel diodes. *Proc. IEEE* **80**, 1810-1826 (1992).
25. Gao, J. R. *et al.* Terahertz superconducting hot electron bolometer heterodyne receivers. *IEEE Trans. Appl. Superconduct.* **17**, 252-258 (2007).
26. Kloosterman, J. L. *et al.* Hot electron bolometer heterodyne receiver with a 4.7-THz quantum cascade laser as a local oscillator. *Appl. Phys. Lett.* **102**, 011123 (2013).
27. Crowe, T. W. *et al*. GaAs Schottky diodes for THz mixing applications. *Proc. IEEE* **80**, 1827-1841 (1992).

28. Berry, C. W. *et al*. Significant performance enhancement in photoconductive terahertz optoelectronics by incorporating plasmonic contact electrodes. *Nat. Commun.* **4**, 1622 (2013).
29. Wang, N. & Jarrahi, M. Noise analysis of photoconductive terahertz detectors. *J. Inf. Mill. THz Waves* **34**, 519-528 (2013).
30. Kerr, A. R. Suggestions for revised definitions of noise quantities, including quantum effects. *Trans. Microwave Theory Tech.* **47**, 325-329 (1999).
31. Hubers, H. W. Terahertz heterodyne receivers. *IEEE J. Sel. Topic. Quantum Electron.* **14**, 378-391 (2008).

**Acknowledgments**
We gratefully acknowledge the financial support from National Aeronautics and Space Administration (NASA) Jet Propulsion Laboratory (JPL) Strategic University Research Partnerships Program. Jarrahi's group gratefully acknowledges the financial support from the Office of Naval Research (contract # N00014-14-1-0573 managed by Dr. Paul Maki) and National Science Foundation (contract #1305931 managed by Dr. Dimitris Pavlidis).

**Author Information**

**Affiliations**
*Electrical and Computer Engineering Department, University of California Los Angeles*
Ning Wang, Semih Cakmakyapan, Yen-Ju Lin & Mona Jarrahi
*Electrical Engineering and Computer Science Department, University of Michigan, Ann Arbor*
Ning Wang & Mona Jarrahi
*Jet Propulsion Laboratory, California Institute of Technology, Pasadena*
Hamid Javadi

**Contributions**
Ning Wang designed and fabricated the device prototypes and performed the heterodyne spectrometer characterization measurements. Semih Cakmakyapan performed the noise temperature measurements. Yen-Ju Lin designed and fabricated the IF circuits. Hamid Javadi assisted with project supervision and heterodyne spectrometer characterization. Mona Jarrahi supervised the project and wrote the manuscript. All authors discussed the results and commented on the manuscript. All data needed to evaluate the conclusions in the paper are present in the paper and the supplementary materials.

**Competing interests**
The authors declare no competing financial interests.

**Corresponding Author**
Correspondence and requests for materials should be addressed to Mona Jarrahi, mjarrahi@ucla.edu.

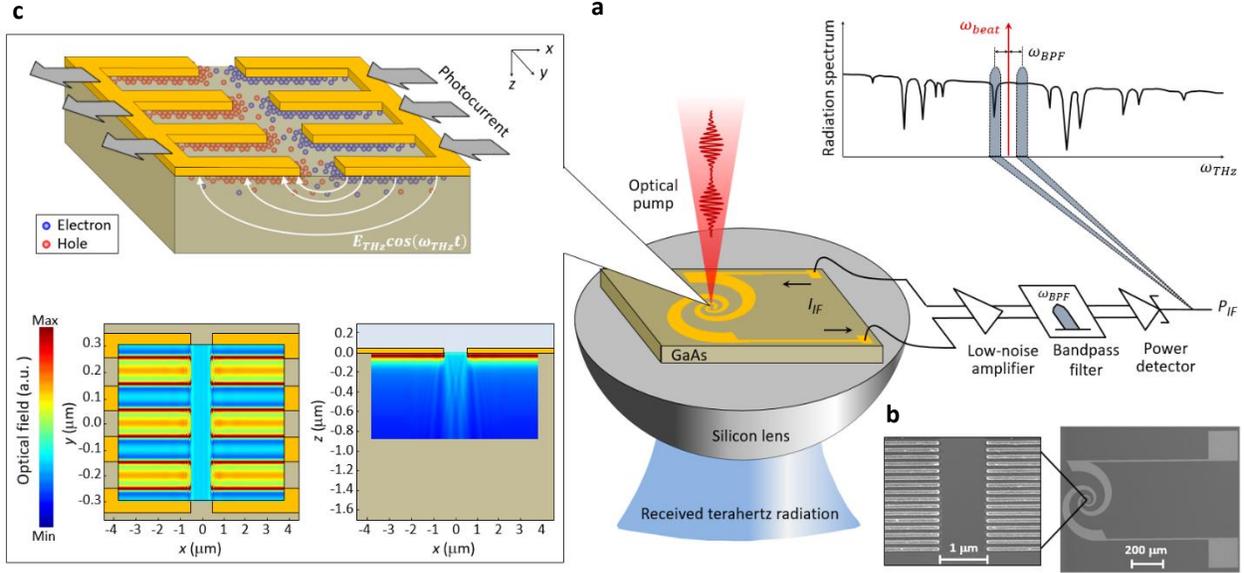

**Figure 1 | Principles of terahertz spectrometry through plasmonic photomixing. a,** When the plasmonic photomixer is pumped by a heterodyning optical beam with a terahertz beat frequency ($\omega_{beat}$), the received terahertz radiation near $\omega_{beat} \pm \omega_{BPF}$ is downconverted to an IF current at $|\omega_{beat} - \omega_{THz}|$, which can be easily detected by RF electronics. **b,** Scanning electron microscopy image of a fabricated plasmonic photomixer prototype. **c**, To achieve high THz-to-RF conversion efficiencies, two nanoscale Ti/Au gratings with a 50 nm thickness, 200 nm pitch, 100 nm spacing, and 300 nm thick $Si_3N_4$ anti-reflection coating are used as the photomixer contacts to enhance the optical pump intensity at the contact-semiconductor interface through the excitation of surface plasmon waves. These plasmonic contact electrode gratings are designed to cover an 8×8 μm$^2$ active area with a tip-to-tip gap of 1 μm. This design of plasmonic gratings provides a high optical power transmission (~ 85%) and a strong plasmonic enhancement when excited by a y-polarized optical pump beam at an ~784 nm wavelength, as illustrated in the top view (*xy*-plane) and cross-sectional view (*xz*-plane) color plots of the optical field in the GaAs substrate (see Methods and Fig. S1).

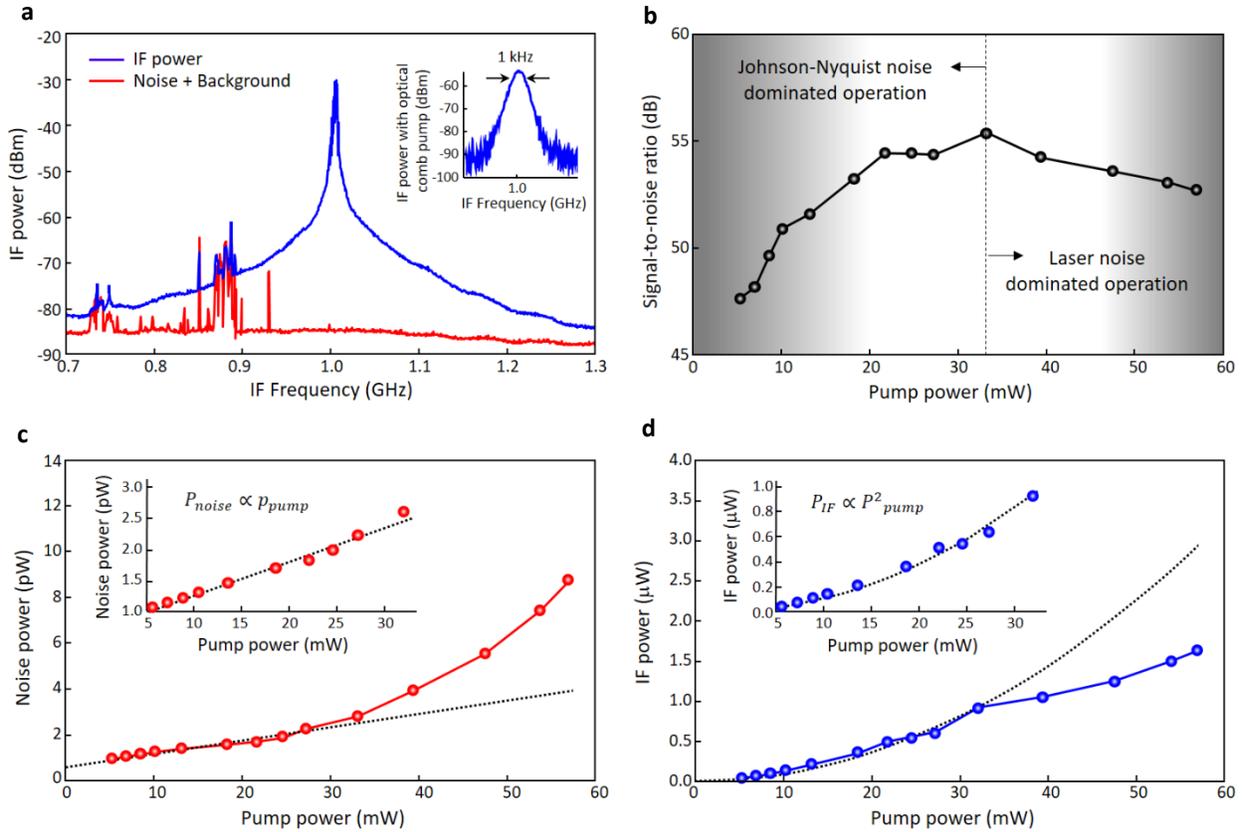

**Figure 2 | THz-to-RF conversion using the fabricated plasmonic photomixer. a,** The downconverted IF spectrum from 0.55 THz to 1 GHz (blue curve) and the IF noise/background in the absence of the 0.55 THz source (red curve) at a 33 mW optical pump power from two DFB lasers. The inset shows the IF spectrum when an optical comb from a Ti:sapphire mode-locked laser is used as the optical pump beam. The measured IF SNR, noise power, and signal power as a function of the optical pump power are shown in **b**, **c**, and **d**, respectively. A tunable attenuator is used to vary the optical pump power in these measurements. The IF noise and signal powers show a linear and quadratic dependence on the optical pump power below 33 mW, as illustrated in **c** and **d** insets with R-squared fits of 98.5% and 99.1%, respectively.

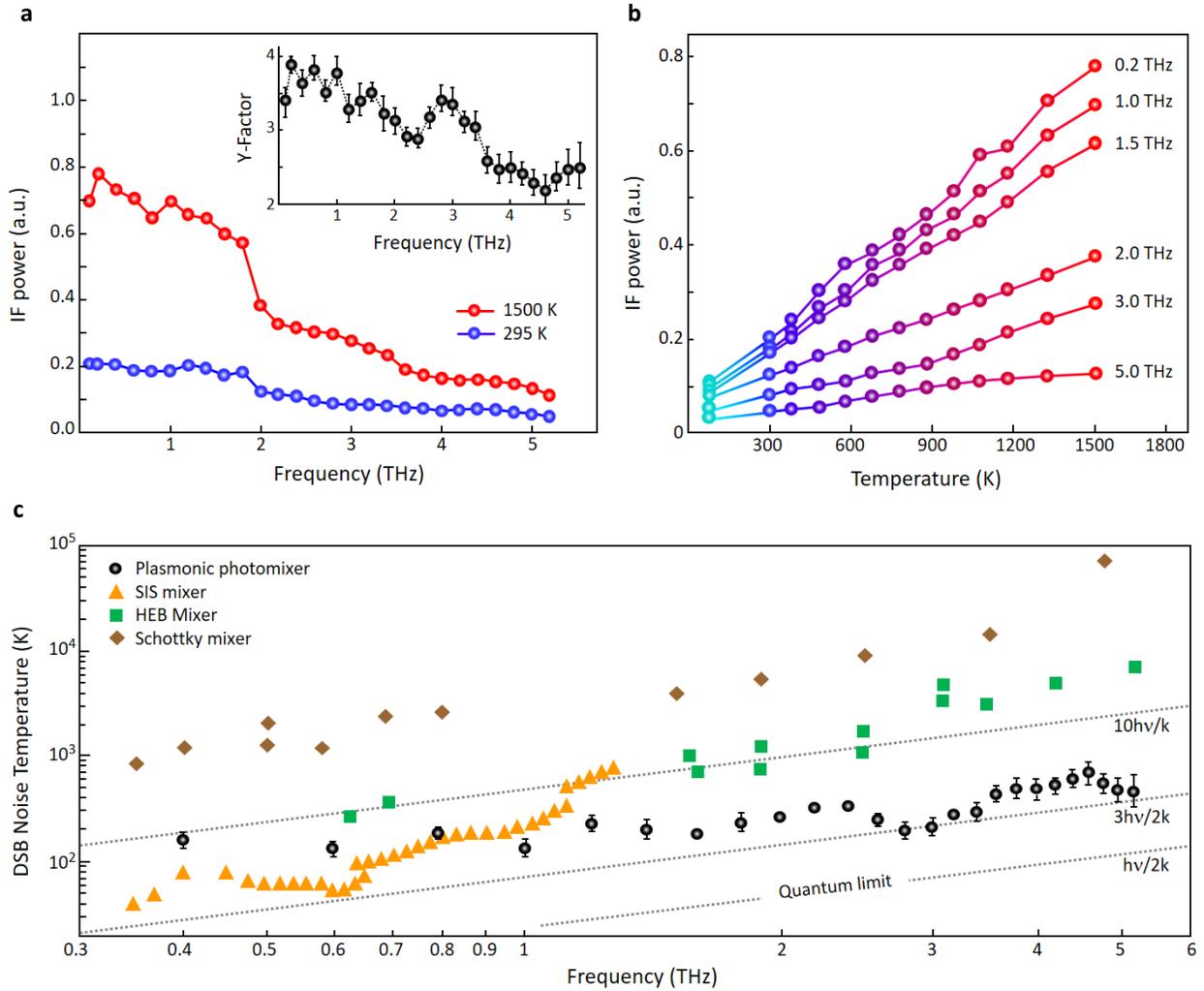

**Figure 3 | Noise temperature characteristics of the fabricated plasmonic photomixer. a,** The measured IF power in response to 1500/295 K hot/cold loads at a 30 mW optical pump power over the 0.1-5 THz frequency range. Inset shows Y-factor values calculated as the measured IF power at 1500 K ($P_{hot}$) divided by the measured IF power at 295 K ($P_{cold}$). **b,** The measured IF power in response to thermal loads in the 77-1500 K range at a 30 mW optical pump power. **c,** DSB noise temperature values of the plasmonic photomixer compared with previously demonstrated Schottky mixers, HEB mixers, and SIS mixers used in conventional spectrometers[31]. The DSB noise temperature values are calculated as $(T_{eff.1500} - Y \cdot T_{eff.295})/(Y-1)$, where $T_{eff.1500}$ and $T_{eff.295}$ are the equivalent temperatures of a blackbody at 1500 K and 295 K, respectively, according to the Callen-Walton definition. The error bars are attributed to the fluctuations in the optical pump power (Fig. S2).

Supplementary Materials for

# Room Temperature Terahertz Spectrometer with Quantum-Level Sensitivity

Ning Wang[1,2] †, Semih Cakmakyapan[1] †, Yen-Ju Lin[1], Hamid Javadi[3], Mona Jarrahi[1]*

[1]Electrical and Computer Engineering Department, University of California Los Angeles, USA.
[2]Electrical Engineering and Computer Science Department, University of Michigan, Ann Arbor, USA.
[3]Jet Propulsion Laboratory, California Institute of Technology, Pasadena, USA.

*Correspondence to: mjarrahi@ucla.edu
†Equal contribution

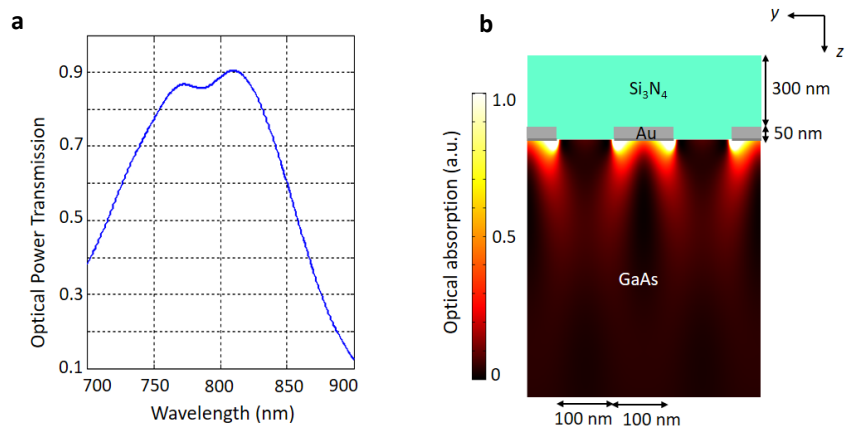

**Figure S1 |** Optical interaction with the designed plasmonic gratings. **a,** Power transmission of a *y*-polarized optical beam through the nanoscale gratings into the LT-GaAs substrate. **b,** Optical power absorption inside the LT-GaAs substrate at a 784 nm wavelength.

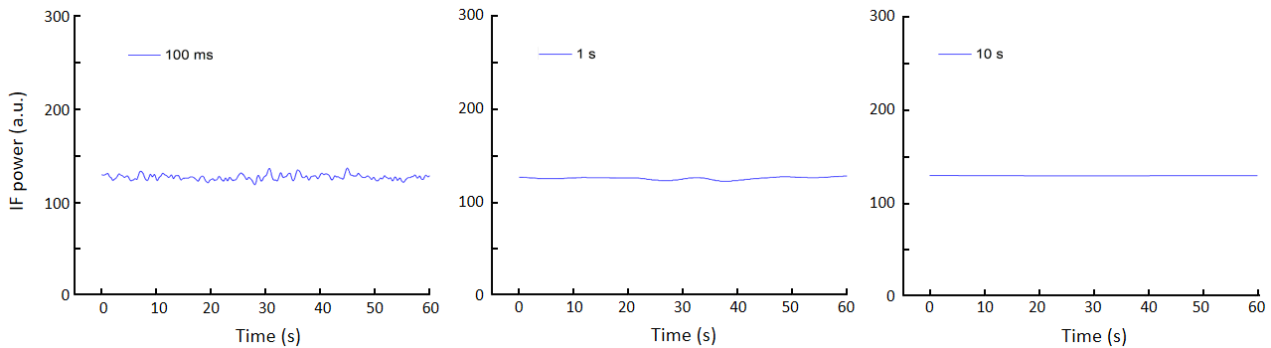

**Figure S2 |** Impact of the integration time on the spectrometer accuracy. The recorded IF power at 0.8 THz over a 60 s timeframe while using a 1500 K load shows a significant reduction in signal fluctuations caused by optical pump power variations when increasing the integration time from 100 ms to 1 s.

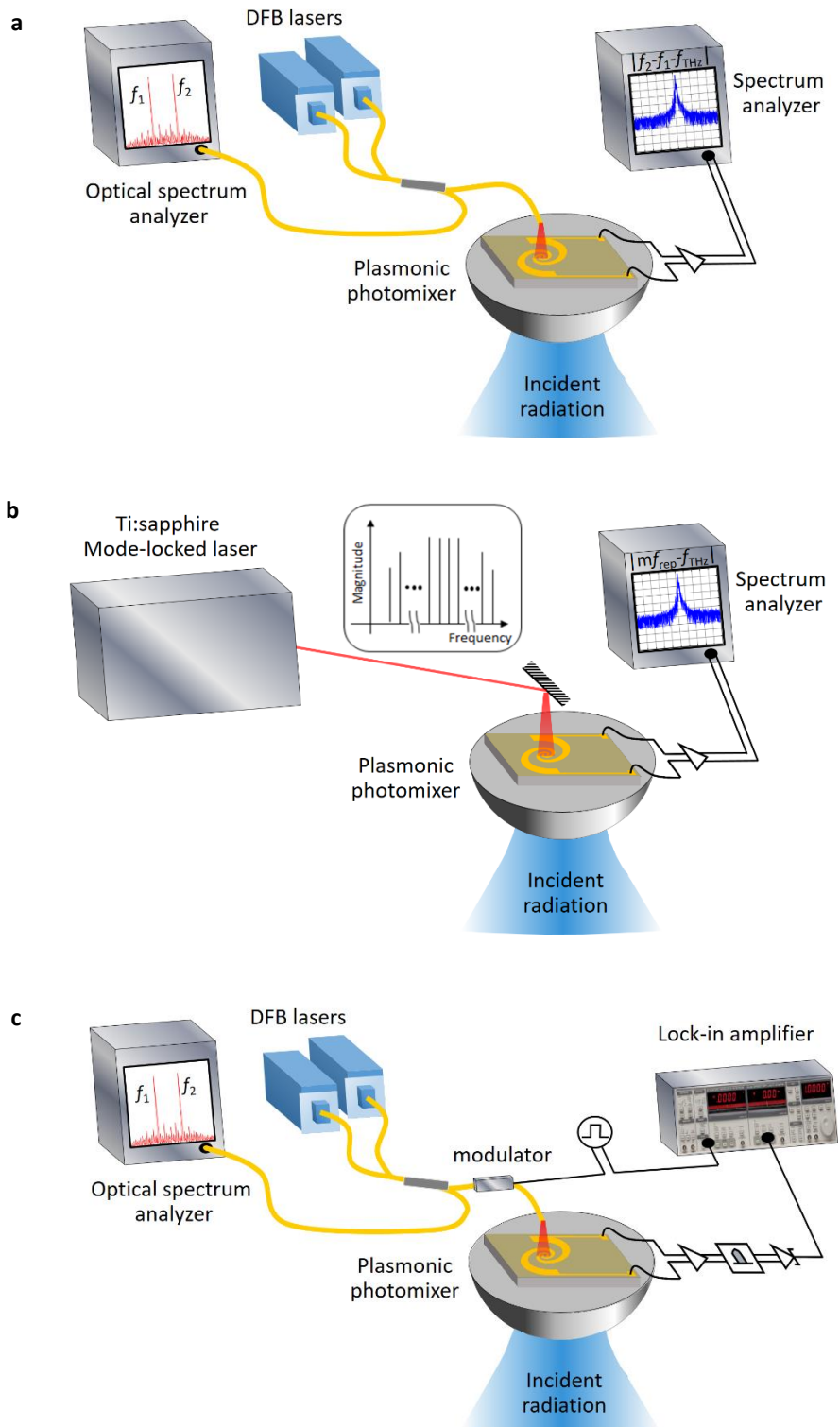

**Figure S3 |** Experimental setup used for **a,** characterizing the operation of the fabricated plasmonic photomixer as a heterodyne terahertz spectrometer, **b,** measuring the linewidth and stability of the terahertz spectrometer, **c,** characterizing the sensitivity of the terahertz spectrometer.

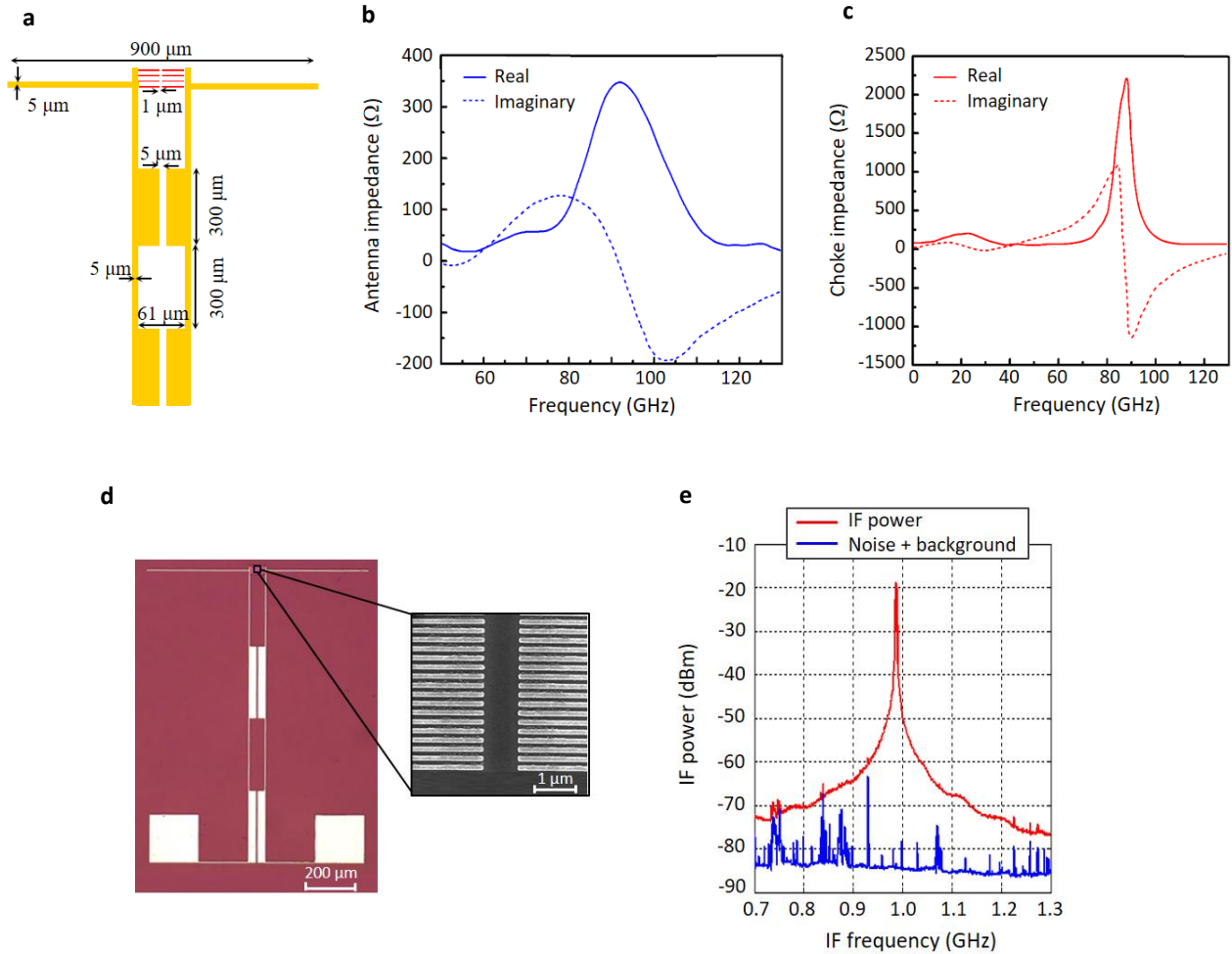

**Figure S4** | Schematic of a polarization-sensitive heterodyne spectrometer, composed of a plasmonic photomixer integrated with a full-wavelength dipole antenna on a LT-GaAs substrate, optimized for operation at ~ 0.1 THz is illustrated in (**a**). Two nanoscale Ti/Au gratings with a 50 nm thickness, 200 nm pitch, 100 nm spacing, and 300 nm thick $Si_3N_4$ anti-reflection coating are used as the photomixer contacts. The plasmonic contact electrode gratings are designed to cover a 30×30 $\mu m^2$ active area with a tip-to-tip gap of 1 µm. A terahertz choke is connected to the plasmonic contact electrodes to route the IF photocurrent out of the plasmonic photomixer. It is made up of four alternating high-impedance and low-impedance quarter-wavelength transmission lines and is designed to provide a high resistance and low reactance loading to the terahertz antenna at ~ 0.1 THz while keeping a 50 Ω impedance at GHz-range IF frequencies. The full-wavelength dipole antenna and terahertz choke are designed using a finite-element-method-based electromagnetic software package (ANSOFT HFSS). Real (solid line) and imaginary (dashed line) parts of the antenna impedance are shown in (**b**), indicating a resonance behavior at ~ 0.1 THz. Real (solid line) and imaginary (dashed line) parts of the choke impedance are shown in (**c**), indicating a high resistance and low reactance loading to the antenna at ~ 0.1 THz while keeping a 50 Ω impedance at a 1 GHz IF frequency. An optical microscope image of a fabricated plasmonic photomixer prototype with a dipole antenna and a scanning electron microscopy (SEM) image of the plasmonic contact electrode gratings are shown in (**d**). Operation of the fabricated plasmonic photomixer as a heterodyne terahertz spectrometer is characterized in response to radiation from a Gunn oscillator (Millitech GDM-10 SN224). To provide the heterodyning optical pump beam, the outputs of two wavelength-tunable, distributed-feedback (DFB) lasers with center wavelengths at 783 nm and 785 nm (TOPTICA #LD-0783-

0080-DFB-1 and #LD-07835-0080-DFB-1) are combined and amplified (Toptica BoosTA Pro) to provide a tunable optical beat frequency at ~ 0.1 THz. The IF output of the plasmonic photomixer is amplified using a low-noise amplifier (Mini-Circuits ZRL-1150) and monitored by an electrical spectrum analyzer. The polarization of the optical pump beam and the Gunn Oscillator radiation are set to be perpendicular to the plasmonic gratings and along the dipole antenna axis, respectively. The observed IF spectrum centered at 1 GHz (red curve) and the noise/background spectrum in the absence of the terahertz radiation (blue curve) are shown in (**e**). The lower spectral peaks observed near 750 MHz and 1.25 GHz are the background radio signals picked up by the IF transmission lines and cables.